\documentclass{article}
\usepackage[utf8]{inputenc}
\usepackage{graphicx}
\usepackage{url}

\title{Reliablocks: Developing Reliability Scores for Optimistic Rollups }
\author{Souradeep Das, Ethan Lam, Varun Vaidya, Sanjay Amirthraj}
\date{}

\begin{document}

\maketitle

\section*{Abstract}
Introducing Reliablocks, an on-chain reliability index for non-finalized blocks in Optimistic Rollups. This was built during the EigenLayer Infinite Hackathon at the Infinite Hacker House at DevCon 2024. As part of this research, we delivered a working Layer AVS WASMI component, a working Eigen Layer AVS component, EigenLayer Solidity smart contracts that work with the AVS component,
a UI dashboard illustrating the reliability score and a derived interest rate for further utilization.

\section*{Introduction}
Optimistic rollups are Layer 2 (L2) scaling solutions for Ethereum that process transactions off-chain and post results on-chain, offering improved scalability and approximately 99\% lower fees, while inheriting Ethereum’s security. These solutions have gained significant traction, with the top two projects, Arbitrum and Optimism, currently holding \$21.2 billion in assets. 

Fast exit providers such as Celer and Across play a crucial role in this ecosystem by fronting funds to users, allowing them to receive funds quicker than Optimism’s standard 7-day finalization period when bridging funds to the Ethereum mainnet. This service comes with an interest charge that compensates providers for the risk they assume.

However, if you are not running a validator, you have no way of determining whether a specific block will be finalized or not. For the general L2 user, it is neither feasible nor expected to run a validator. Thus, they have no reliable way to identify whether a transaction they were part of was valid (inherited validity if it is on top of a valid state) and that the transaction will not be rolled back during the next seven days. This raises an important question: \textit{How long should a user on an L2 wait before they can deem their transaction final?}

\section*{Reliablocks}
We are building an on-chain reliability index for non-finalized blocks in Optimistic Rollups to help non-validators estimate the likelihood of a non-finalized block being finalized. Such an index would provide valuable information to users without requiring them to trust a specific entity or validator for the information. A dynamic score, instead of a binary final/non-final answer, also assigns a weight to the reliability of what multiple agents think about the finality of a specific block.

\section*{AVS Integration for Reliability Score}

One of our (initial) key metrics for determining the reliability score is data collected during fast exits. Actively Validated Services (AVS) taps into Ethereum's validators to validate data collected, making it feasible to translate fast exit data into a reliability score. 

Every L2 transaction is batched and deposited onto Layer 1 (L1). This means that all fast exit transaction data is available on L1 and can be easily verified on-chain. Since the data is part of the calldata, it is verifiable on-chain, but not retrievable by an external contract. The counterpart of the fast exit on L2 is a liquidity transfer on an L1. While both of these are verifiable on-chain, they need to be inputted by an operator, whom we must blindly trust. This is where AVS comes into play, helping set up a validation mechanism where operators provide data; good operators receive rewards while malicious operators are slashed for providing fraudulent information.

Rather than simply providing a binary final/non-final answer for each L2 block, the validator assigns a numerical value, quantifying the finality they are proposing. This approach creates an effective way for end users to trust multiple operators at once by referring to a single metric, while also giving the conditions of trust to the users' discretion.

Overall, AVS supports this research project by offering several key features:
\begin{itemize}
    \item Slashing a validator who provides incorrect data, utilizing restaked security guarantees.
    \item Making the reliability index available on-chain, so it can be further utilized by other cross-chain bridges, smart contract logic, and more. (For instance, users can be descriptive with their risk indices when interacting with a contract.)
    \item The on-chain reliability index can also be deposited onto the L2, where it will relay the index data back to L2 decentralized applications (dApps), extending its application to apps running on L2.
\end{itemize}
See \textbf{Figure 2} for a visual image of fast exit reliability metric determination.

An on-chain reliability index for non-finalized blocks in Optimistic Rollups would greatly enhance user experience by reducing uncertainty and improving risk assessment for transactions. It would provide valuable insights for users and fast exit providers, enabling informed decisions regarding withdrawals and fund fronting. This increased transparency could boost confidence in Optimistic Rollups, facilitating better interoperability solutions and greater adoption of Layer 2 solutions on Ethereum.

\section*{Introducing Our Score}
\subsection*{Quantifying L2 Block Finality}
How can the finality of an L2 block be quantified? Every L2 has certain actors who actively make decisions based on the validity of a state, even to the extent of putting money at risk to support their decisions. These actors are assumed to run a full validator. Who are they? They include fast exit providers and liquidity pool providers who take someone's position on the L2 themselves, ensuring the validity of the rollup's state while providing fast liquidity in return.

\subsection*{Dynamic Reliability Score}
Our score is a metric to determine the likelihood of block validity, using the number and value of fast exits in an L2 block as a key determinant. Blocks with more fast exits, especially high-value ones, are considered to be more reliable, as they indicate validators' confidence in the block's validity and confirm the changed state of the block. Since each block is assigned a financial weight based on the fast-exit providers enabling exits on it, our score becomes dynamic. Every subsequent fast-exit provider effectively attests to the validity of the previous state on which they are providing the service. Over time, a block's reliability score increases, akin to block confirmations on L1. These cumulative scores serve as a suitable indicator for entities, such as exchanges, to wait for a specific reliability score before accepting a transaction on an L2.

\subsection*{Integration into Cross-Chain Products}
To integrate the score into cross-chain products, we use it to determine interest rates. Blocks with higher reliability scores are associated with lower interest rates, reflecting a lower perceived risk. We modeled this relationship using a negative exponential decay function to assign a score between 0 and 100, normalized for each block. Based on this normalization, interest rates are expected to range between 1\% and 3\%, depending on the score.
\section*{Architecture}
See Figure 1

\section*{User Interface}
See Figure 3

\section*{Primary Stakeholders}
The primary stakeholders in this system include:
\begin{enumerate}
    \item Optimistic rollup users who want to ascertain the finality of their transactions.
    \item Optimistic rollup users seeking faster withdrawals.
    \item Centralized exchanges (CEXs) on L2 requiring a reliable indicator (confirmations) to accept certain deposit transactions.
    \item Fast exit providers who validate transactions and assume financial risk.
    \item Rollup operators and validators responsible for processing transactions.
    \item The proposed reliability scoring system, which could evolve into a fast exit provider itself.
\end{enumerate}

\section*{Future Steps}
We plan to research and develop a stronger algorithm with other weighted indicators that provide a clearer understanding of the chances of a blocks' finalization. For this to become a business, we look to create an insurance protocol based on our metrics. Since we are solving the "How trustworthy are these validators?" problem, we need to find a way to keep trust in our system (without requiring blind trust from our users) while keeping a competitive advantage within the ecosystem. Our proposed scoring system addresses this concern by offering a transparent and dynamic measure of reliability, reducing the need for blind trust in specific entities, but may require further tweaks to keep this balance.

\section*{Hackathon Development and Feedback: 11/10-11/15}

\subsection*{1. Setup (11/10-11/11)}
\begin{itemize}
    \item Learned to build on Layer (new platform for us).
    \item Successfully ran the square and oracle example SDK provided by Layer.
\end{itemize}

Walking through the Layer docs to do basic tasks was easy — we were able to obtain funds from the faucet, deploy contracts, and run the tests for the square and oracle examples.

\subsection*{2. Challenges with Rust Smart Contracts on Layer’s Chain}
\subsubsection*{2.1 Fetching Issues}
\begin{itemize}
    \item Unable to interact with smart contracts on Layer.
    \item Difficulties in querying contracts.
\end{itemize}

We encountered issues when we began trying to tie our platform together. 

\subsubsection*{2.2 WASI Component Setup}
\begin{itemize}
    \item Successfully set up WASI components.
    \item Implemented result returning and wasmatic test components.
\end{itemize}

We built the off-chain WASI component that would allow operators to submit tracked fast exit transactions from Optimism. The WASM component takes a L2 block number as input and returns the reliability score as a response. We were able to run tests for these WASI components we developed. The component was deployed to the Task Queue contract, and the wasmatic run/test commands were executed successfully to test the request responses to our WASM.

\subsection*{3. Encountered Layer-Specific Challenges (11/14-11/15)}
\subsubsection*{3.1 Task Queue Problems}
\begin{itemize}
    \item Empty task queue when attempting to read records/tasks.
    \item Successful CLI task running and testing, but unclear on-chain response confirmation.
\end{itemize}

\subsubsection*{3.2 Documentation Gaps}
\begin{itemize}
    \item Lack of advanced contract CLI command documentation.
    \item Insufficient information on actual contract deployment interaction.
    \item Limited resources on Layer SDK and underlying Rust blockchain.
\end{itemize}

\subsection*{4. Contract Interaction Hurdles}
\subsubsection*{4.1 Contract Querying}
\begin{itemize}
    \item Lack of clarity on the "msg" option format in CLI for contract queries.
\end{itemize}

\subsubsection*{4.2 Storage Layout}
\begin{itemize}
    \item Unclear contract storage layout, especially for default and result storage.
\end{itemize}

\subsection*{5. Alternative AVS Development}
\begin{itemize}
    \item Successfully built an alternate AVS using Layer.
    \item Had trouble connecting to the frontend.
\end{itemize}

Worked around most of the above Layer-based challenges, but had integration problems. Decided to build an EigenLayer AVS as a backup.
\subsection*{6. Parallel Development on EigenLayer}
\begin{itemize}
    \item Built AVS on EigenLayer as well.
    \item Achieved a working iteration in 3-4 hours (compared to 3 days on Layer).
\end{itemize}
The Layer AVS worked on Terminal. We integrated the EigenLayer AVS.
\subsection*{7. Identified Areas for Improvement}
\begin{itemize}
    \item More comprehensive documentation needed for Layer.
    \item Better resources required for Cosmos and Rust blockchain interactions.
    \item Need for more robust contract interaction tools and documentation.
\end{itemize}
Our team had a lot more experience with EVM rather than CosmWASM.

\section*{Citations}
\begin{enumerate}
    \item L2Beat. "TVL in Optimistic Rollups". Available at: \url{https://l2beat.com/scaling/tvl}
    \item Aft 2023. "LIPIcs: AFT.2023.22". Available at: \url{https://drops.dagstuhl.de/storage/00lipics/lipics-vol282-aft2023/LIPIcs.AFT.2023.22/LIPIcs.AFT.2023.22.pdf}
    \item Ethereum.org. "Optimistic Rollups". Available at: \url{https://ethereum.org/en/developers/docs/scaling/optimistic-rollups/}
    \item Alchemy. "Overview of Optimistic Rollups". Available at: \url{https://www.alchemy.com/overviews/optimistic-rollups}
    \item OSIZ Technologies. "Optimistic Rollups Development". Available at: \url{https://www.osiztechnologies.com/blog/optimistic-rollups-development}
\end{enumerate}

\begin{figure}[h!]
    \centering
    \includegraphics[width=\textwidth]{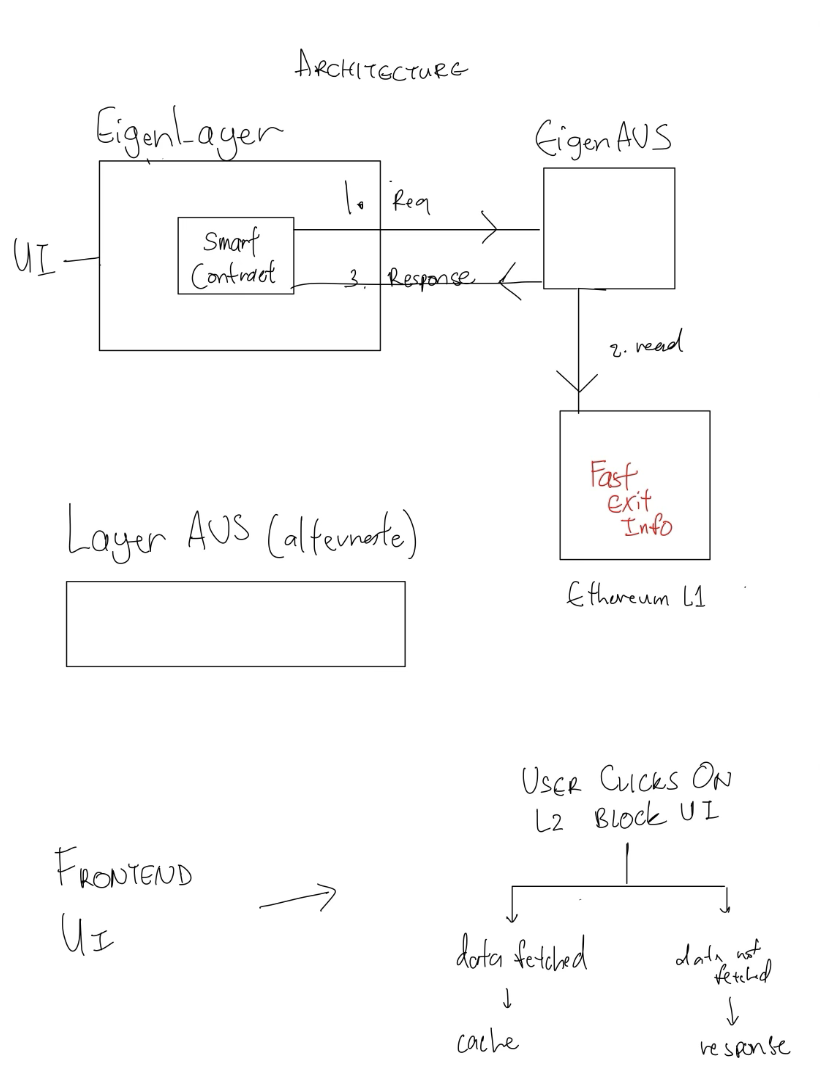}
    \caption{hand-drawn diagram of ReliaBlocks architecture}
    \label{fig:image-label}
\end{figure}
\begin{figure}[h!]
    \centering
    \includegraphics[width=\textwidth]{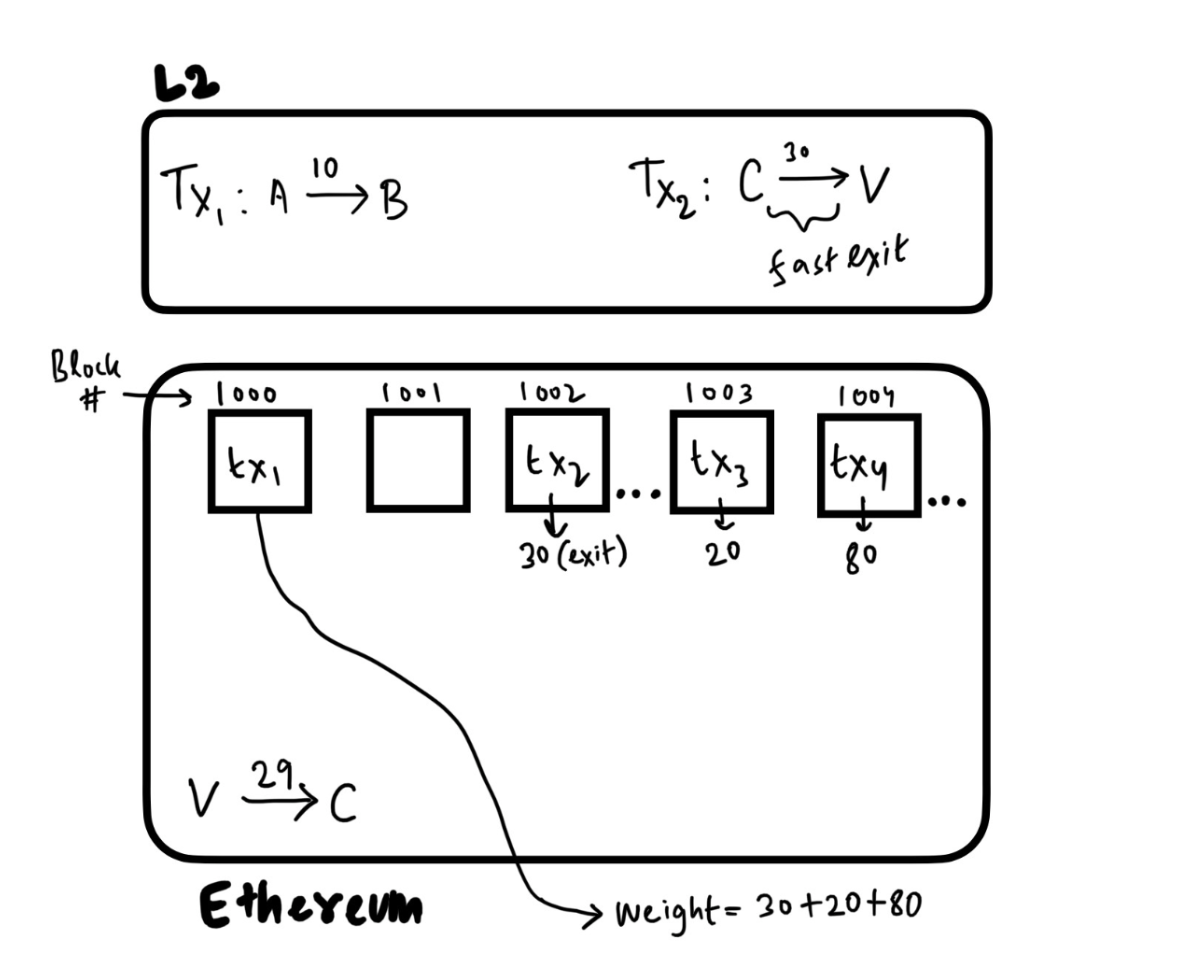}
    \caption{Visual Description of Fast Exit Reliability Definition}
    \label{fig:image-label}
\end{figure}
\begin{figure}[h!]
    \centering
    \includegraphics[width=\textwidth]{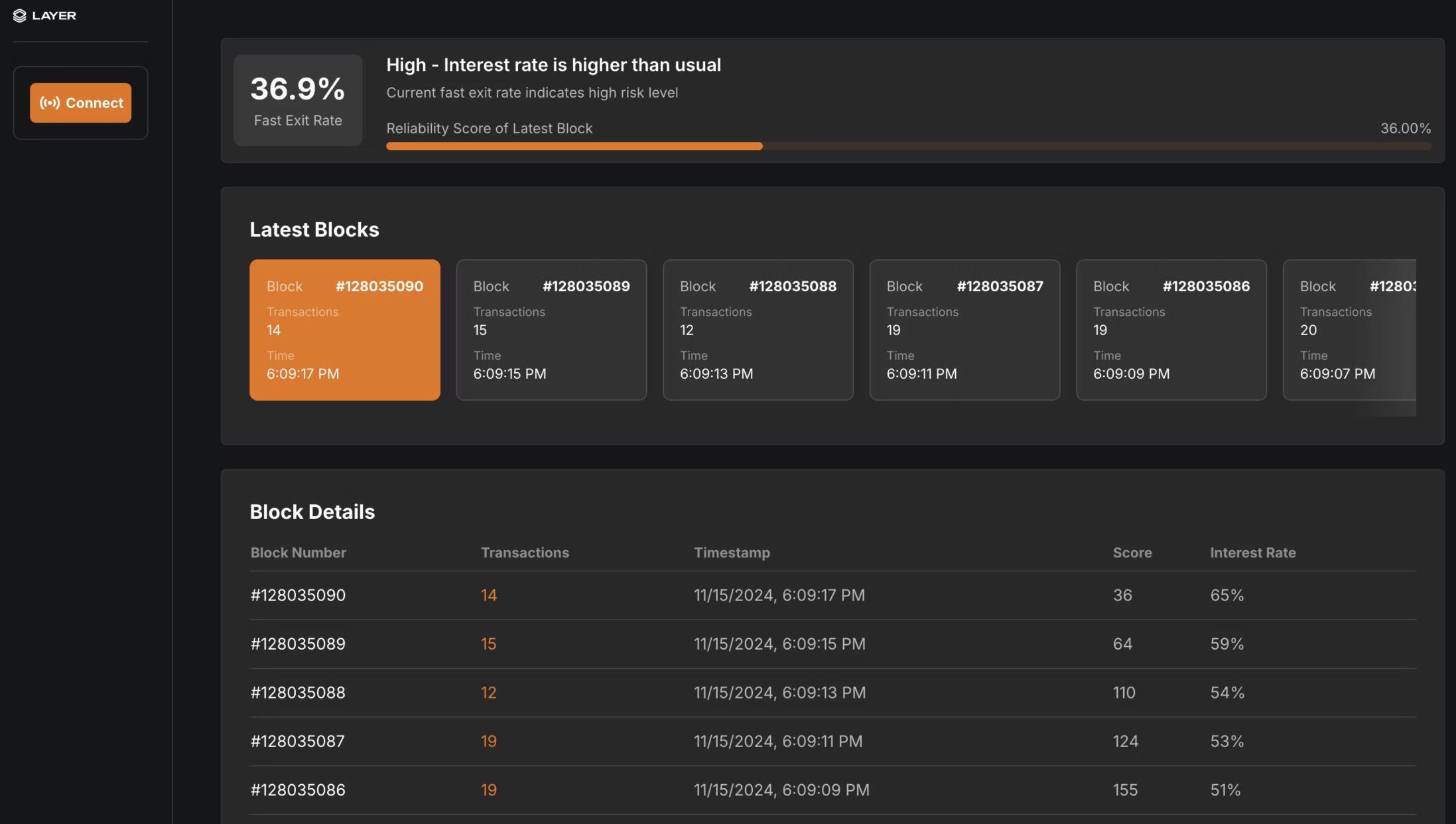}
    \caption{User Interface}
    \label{fig:image-label}
\end{figure}

\end{document}